\documentclass[aps,prc,superscriptaddress,twoside,twocolumn,nofootinbib,10pt,%
showpacs,floatfix]{revtex4-1}

\usepackage{amsmath,amssymb}
\usepackage{graphicx,bm}
\usepackage{epstopdf}
\usepackage{ulem} 
\usepackage[usenames]{color}
\usepackage{float}

\allowdisplaybreaks

\newcommand{\Tr}{\textrm{Tr}}

\newcommand{\vsla}{v\hspace{-.47em}/}

\newcommand{\olra}{\overleftrightarrow}
\newcommand{\fpi}{f_{\pi}}

\newcommand{\Lamc}{\Lambda_c}
\newcommand{\Sigc}{\Sigma_c}
\newcommand{\Sigcs}{\Sigma_c^*}
\newcommand{\Db}{\bar{D}}
\newcommand{\Dbs}{\bar{D}^*}

\newcommand{\non}{\nonumber}
\newcommand{\Jp}{J/\psi}
\newcommand{\del}{\partial}

\begin{document}

\title{Hidden Charm Pentaquark $P_c(4380)$ and Doubly Charmed Baryon $\Xi_{cc}^*(4380)$ as Hadronic Molecule States}
%%%%%%%%%%%%%%%%%%%%%%%%%%%%%%%%%%%%%%%%%%%%%%

\author{Yuki Shimizu}
\email{yshimizu@hken.phys.nagoya-u.ac.jp}
\affiliation{Department of Physics,  Nagoya University, Nagoya, 464-8602, Japan}

\author{Masayasu Harada}
\email{harada@hken.phys.nagoya-u.ac.jp}
\affiliation{Department of Physics,  Nagoya University, Nagoya, 464-8602, Japan}

\date{\today}

%%%%%%%%%%%%%%%%%%%%%%%%%%%%%%%

\begin{abstract}
We study hadronic molecular states in a coupled system of $\Jp N - \Lamc\Db^{(*)} - \Sigc^{(*)}\Db^{(*)}$ in $I(J^P) = \frac{1}{2}(\frac{3}{2}^-)$ channel, 
using the complex scaling method combined with the Gaussian expansion method. 
We construct the potential  including one pion exchange and one $D^{(*)}$ meson exchange with $S$-wave orbital angular momentum.
We find that the both mass and width of the pentaquark $P_c(4380)$ can be 
reproduced 
within a reasonable parameter region, and that its main decay mode is $\Lamc\Dbs$.
We extend our analysis to a coupled system of $\Lamc D^{(*)} - \Sigc^{(*)}D^{(*)}$ in $I(J^P) = \frac{1}{2}(\frac{3}{2}^-)$ channel. We find that there exists a doubly charmed baryon of $ccqq\bar{q}$ type as a hadronic molecule, the mass and width of which are quite close to those of $P_c(4380)$.
\end{abstract}
%%%%%%%%%%%%%%%%%%%%%%%%%%%%%%%%%%%%%%
%%%%%%%%%%%%%%%%%%%%%%%%%%%%%%%%%%%%%%

\maketitle

%%%%%%%%%%%%%%%%%%%%%%%%%%%%%%%%%%%%%%%%%%%%%%%%%%%%%%%

\section{Introduction}
\label{sec:Intro}
In 2015, the LHCb experiment announced the observation of the hidden charm pentaquark $P_c(4380)$ and $P_c(4450)$~\cite{Aaij:2015tga,Aaij:2016phn,Aaij:2016ymb}.
The mass and width of $P_c(4380)$ are $M=4380\pm8\pm29$MeV and $\Gamma=205\pm18\pm86$ MeV
and  those  of $P_c(4450)$ are $M = 4449.8\pm1.7\pm2.5$ and $\Gamma = 39\pm5\pm19$MeV.
Their spins and parities are not well determined; most likely $J^P=(3/2^- , 5/2^+)$.

Some theoretical works were done before the LHCb result in 
Refs.~\cite{Wu:2010vk, Yang:2011wz, Wang:2011rga, Wu:2012md, Uchino:2015uha}.
After the LHCb announcement, there are many theoretical analyses based on 
the hadronic molecule picture 
\cite{Chen:2015loa, He:2015cea, Chen:2015moa, Huang:2015uda, Roca:2015dva, Meissner:2015mza, Xiao:2015fia, Burns:2015dwa, Kahana:2015tkb, Chen:2016heh, Chen:2016otp, Shimizu:2016rrd, Yamaguchi:2016ote, He:2016pfa, Ortega:2016syt, Azizi:2016dhy, Geng:2017hxc}, 
diquark-diquark-antiquark (diquark-triquark) picture
\cite{Maiani:2015vwa, Lebed:2015tna, Anisovich:2015cia, Li:2015gta, Wang:2015epa, Zhu:2015bba},
compact pentaquark states
\cite{Santopinto:2016pkp, Takeuchi:2016ejt, Wu:2017weo},
and triangle singularities
\cite{Guo:2015umn,Liu:2015fea,Mikhasenko:2015vca,Liu:2016dli,Guo:2016bkl,Bayar:2016ftu}.
The decay behaviors are studied in Refs.\cite{Wang:2015qlf, Lu:2016nnt, Shen:2016tzq, Lin:2017mtz}.

In Ref.~\cite{Shimizu:2016rrd}, effect of $\Sigma_c^*\bar{D}-\Sigma_c\bar{D}^*$ coupled channel is studied in the hadronic molecule picture for  the hidden charm pentaquark with $I(J^P) = 1/2(3/2^-)$, 
by using the one-pion exchange potential with $S$-wave orbital angular momentum.
It was shown that  there exists a bound state with 
the binding energy of several MeV below $\Sigma_c^\ast\bar{D}$ threshold,  which 
 is mainly made from a $\Sigma_c^\ast$ and a $\bar{D}$.
In Ref.~\cite{Yamaguchi:2016ote},
the coupled channel effect to $\Lambda_c\bar{D}^{(*)}$ was shown to be important to investigate the $P_c$ pentaquarks.
In Ref.~\cite{Lin:2017mtz}, 
decay behaviors of hadronic molecule states of $\Sigma_c^*\bar{D}$ and $\Sigma_c\bar{D}^*$  to $J/\psi N$ are studied and it was shown that  the contribution of $J/\psi N$ 
is small for the $P_c(4380)$ as the $\Sigma_c^*\bar{D}$ molecule.
However, in our best knowledge, study of the effect of $J/\psi N$ in full coupled channel analysis, which reproduce both the mass and width of $P_c(4380)$, was not done so far.

In this paper, we make a coupled channel analysis including $\Jp N$ in addition to $\Lambda_c\bar{D}^{(*)} - \Sigma_c^{(*)}\bar{D}^{(*)}$ with $S$-wave orbital angular momentum.
Here we construct a relevant potential from exchange of one pion and $D^{(\ast)}$ mesons.
Our result shows that both the mass and width of $P_c(4380)$ are within experimental errors for reasonable parameter region, and that the effect from 
$J/\psi N$ channel is very small.
In other word, the observed mass and width of $P_c(4380)$ are well reproduced dominantly by one-pion exchange potential for $\Lambda_c\bar{D}^{(*)} - \Sigma_c^{(*)}\bar{D}^{(*)}$ coupled channel.

Since the one-pion exchange potential for $\Lambda_c{D}^{(*)} - \Sigma_c^{(*)}{D}^{(*)}$ coupled channel is same as the one for $\Lambda_c\bar{D}^{(*)} - \Sigma_c^{(*)}\bar{D}^{(*)}$ coupled channel,
we expect the existence of a doubly charmed baryon with $I(J^P) = \frac{1}{2}\left(\frac{3}{2}^-\right)$ having the mass and width close to those of $P_c(4380)$, which we call $\Xi_{cc}^*(4380)$.
In the latter half of this paper, we demonstrate that $\Xi_{cc}^*(4380)$ does exist in our model, which actually has the mass and width quite close to those of $P_c(4380)$.

This paper is organized as follows:
In Sec.~\ref{sec:pot}, we show the potentials which we use in our analysis.
We study the pentaquark $P_c(4380)$ in Sec.~\ref{sec:pc4380}, and the doubly charmed baryon $\Xi_{cc}^*(4380)$ in Sec.~\ref{sec:xicc4380}.  Finally, we will give a brief summary and discussions 
in Sec.~\ref{sec:summary}.

%%%%%%%%%%%%%%%%%%%%%%%%%%%%%%%%%%%%%%%%%%%%%%%%%

\section{Potential}
\label{sec:pot}

In this section, 
we construct 
a potential for our coupled channel analysis based on the heavy quark symmetry and the chiral symmetry.
We include one-pion exchange contribution for 
$\Lambda_c\bar{D}^{(*)} - \Sigma_c^{(*)}\bar{D}^{(*)}$ coupled channel and $D^{(*)}$ meson exchange for adding $J/\psi N$ channel.

For constructing effective interactions of $D$ and $D^\ast$ mesons, it is convenient to use the following heavy meson field $H$ defined as ~\cite{Falk:1991nq, Wise:1992hn, Cho:1992gg, Yan:1992gz}
\begin{align}
H &= \frac{1+\vsla}{2}\left[ D_{\mu}^{*}\gamma^{\mu} + iD\gamma_{5} \right]\ , \label{def H}\\
\bar{H} &= \gamma_0 H^{\dagger} \gamma_0~.
\end{align}
where $D$ and $D^\ast$ are the pseudoscalar and vector meson fields, respectively, and
$v$ denotes the velocity of the heavy mesons.

The pion field is introduced by the spontaneous chiral symmetry breaking $\textrm{SU}(2)_{\textrm{R}}\times\textrm{SU}(2)_{\textrm{L}} \to \textrm{SU}(2)_{\textrm{V}}$.
The fundamental quantity is 
\begin{align}
A_{\mu} = \frac{i}{2}\left( \xi^{\dagger}\del_{\mu}\xi - \xi\del_{\mu}\xi^{\dagger} \right)~,
\end{align}
where $\xi = \exp(i\hat{\pi}/\sqrt{2}\fpi)$.
The pion decay constant is $\fpi \sim 92.4$MeV and the pion field $\hat{\pi}$ is defined by a $2 \times 2$ matrix 
\begin{align}
\hat{\pi} = \left(
\begin{array}{cc}
\pi^0/\sqrt{2} & \pi^+ \\
\pi^- & -\pi^0/\sqrt{2}
\end{array}
\right)~.
\end{align}

The interaction Lagrangian for  the 
heavy meson and pions with least derivatives \cite{Wise:1992hn,Yan:1992gz, Cho:1992gg}
 is given by 
\begin{align}
\mathcal{L}_{HH\pi} &= g\Tr\left[ \bar{H}H\gamma_{\mu}\gamma_5A^{\mu} \right]~,
\end{align}
where $g$ is a
dimensionless coupling constant.
The explicit interaction terms can be written as
\begin{align}
\mathcal{L}_{D^*D^*\pi} &= \frac{\sqrt{2}ig}{\fpi}\epsilon^{\mu\nu\rho\sigma}\bar{D}_{\mu}^*D_{\nu}^*\del_{\rho}\hat{\pi} v_{\sigma}~, \\
\mathcal{L}_{D^*D\pi} &= \frac{\sqrt{2}ig}{\fpi} \left( \bar{D}_{\mu}^*D\del^{\mu}\hat{\pi} - \bar{D}D_{\mu}^*\del^{\mu}\hat{\pi} \right)~,
\end{align}
by expanding the $A_{\mu}$ and $H$ fields.
Note that the 
$DD\pi$ interaction term is prohibited by the parity invariance.
The coupling constant 
$g$ is determined as $|g| = 0.59$ from the decay of $D^* \to D\pi$ \cite{Olive:2016xmw}.
The sign of $g$ cannot be decided by the above decay, however we use $g = 0.59$ in the following analysis.

For introducing $\Sigma_c$ and $\Sigma_c^\ast$, we
define the following superfield $S_{\mu}$ for $\Sigma_c$ and $\Sigma_c^*$~\cite{Liu:2011xc}:
\begin{align}
S_{\mu} &= \Sigma_{c\mu}^* - \sqrt{\frac{1}{3}}\left( \gamma_{\mu} + v_{\mu} \right)\gamma_5\Sigma_c~,
\label{eq:superfield}
\end{align}
where the single heavy baryon fields $\Lambda_c$ and $\Sigma_c$ are expressed by the $2 \times 2$ matrices as
\begin{align}
\Lambda_c &= \left(
    \begin{array}{cc}
      0 & \Lambda_{c}^{+} \\
      -\Lambda_{c}^{+} & 0 
    \end{array}
\right) , \quad \Sigma_c = \left(
    \begin{array}{cc}
      \Sigma_{c}^{++} & \frac{1}{\sqrt{2}}\Sigma_{c}^{+} \\
      \frac{1}{\sqrt{2}}\Sigma_{c}^{+} & \Sigma_{c}^{0} \\
    \end{array}
\right)~,
\end{align}
and the matrix field of $\Sigma_c^*$ is defined similarly to the $\Sigma_c$.
The interaction Lagrangian for  the 
heavy baryon and pions is given by~\cite{Falk:1991nq, Yan:1992gz} 
\begin{align}
\mathcal{L}_{BB\pi} &= \frac{3ig_1}{2}v_{\sigma}\epsilon^{\mu\nu\rho\sigma}\Tr\left[ \bar{S}_{\mu}A_{\nu}S_{\rho} \right] \non \\
&\hspace{8mm} + g_4\Tr\left[ \bar{S}^{\mu}A_{\mu}\Lambda_c \right] + H.c.~,
\end{align}
where $g_1$ and $g_4$ are dimensionless coupling constants.
We use $g_4=0.999$ determined from the $\Sigma_c^* \to \Lambda_c \pi $ decay. 
The value of 
$g_1$ cannot be determined by experimental decay, so we use $g_1=\frac{\sqrt{8}}{3}g_4 = 0.942$ estimated by the quark model in Ref.~\cite{Liu:2011xc} as a reference value, and vary its value about 20\%, $0.753$-$1.13$.

We include $J/\psi$ together with $\eta_c$ using a $\bar{c}c$
spin doublet field $\mathcal{J}$ as~\cite{Jenkins:1992nb, Wang:2015xsa}
\begin{align}
\mathcal{J} &= \frac{1+\vsla}{2}\left( \left( J/\psi\right)^{\mu}\gamma_{\mu} - \eta_c\gamma_5 \right)\frac{1-\vsla}{2}~.
\end{align}
In the following analysis, we use only $\left(J/\psi\right)^{\mu}$ field.
The interaction of ${\mathcal J}$ to the heavy mesons $D^{(*)}$ and its anti-particles $\bar{D}^{(*)}$ is expressed as ~\cite{Wang:2015xsa}
\begin{align}
\mathcal{L}_{\mathcal{J}HH} &= G_1\Tr\left[ \mathcal{J}\bar{H}_A\olra{\del}_{\mu}\gamma^{\mu}\bar{H} + H.c. \right]~,
\label{eq:JHH}
\end{align}
where $\olra{\del}_{\mu} = \overrightarrow{\del}_{\mu} - \overleftarrow{\del}_{\mu}$.
The field $H$ is defined in Eq.~(\ref{def H}), and its anti-particle field $H_A$ is defined as 
\begin{align}
H_A &= \left[ \bar{D}_{\mu}^{*}\gamma^{\mu} + i\bar{D}\gamma_{5} \right] \frac{1-\vsla}{2} \ .
\end{align}
We estimate the value of the coupling constant  $G_1$ by comparing  it  with  the  $\phi KK$ coupling.
Regarding the strange hadrons as heavy hadrons, we can write the effective Lagrangian for $\phi K \bar{K}$ in the same form as the one in Eq.~(\ref{eq:JHH}).
Using the value of $\phi K \bar{K}$ coupling $G_{1(\phi KK)}$ determined from the $\phi \to K \bar{K}$ decay: $G_{1(\phi KK)}=4.48[\textrm{GeV}^{-3/2}]$,
we estimate the value of $G_1$ as
\begin{align}
G_1 = G_{1(\phi KK)}\sqrt{\frac{m_{\phi}m_K^2}{m_{\Jp}m_D^2}} = 0.679[\textrm{GeV}^{-3/2}].
\end{align}

The Lagrangian for the interactions among single heavy baryons, $D^{(*)}$ mesons and nucleons is 
given by
\begin{align}
\mathcal{L}_{BHN} &= 
G_2 \left(\tau_2\bar{S}_{\mu}\right)H\gamma_5\gamma^{\mu}N + H.c. \non \\
&+ G_3 \left(\tau_2\bar{\Lambda}_{c}\right)HN + H.c. ~.
\end{align}
We estimate the values of $G_2$ and $G_3$ using $g_{\Sigma_c DN}=2.69$ and $g_{\Lambda_c DN}=13.5$~\cite{Lu:2016nnt, Lin:2017mtz, Garzon:2015zva, Liu:2001ce}.
Considering the differences of the normalization of a heavy meson field by
$\sqrt{m_D}$, we estimate them as
\begin{align}
G_2 &= - \frac{g_{\Sigma_c DN}}{\sqrt{3m_D}} = -1.14[\textrm{GeV}^{-1/2}]~, \\
G_3 &= \frac{g_{\Lambda_c DN}}{\sqrt{m_D}} = 9.88[\textrm{GeV}^{-1/2}]~.
\end{align}
Here the factor 
$-\frac{1}{\sqrt{3}}$ comes from the coefficient in Eq.~(\ref{eq:superfield}).
The estimations of 
the values of  $G_{1,2,3}$ are very rough.
We will discuss the effects of ambiguities in the folllowing sections.

We constract the one-pion exchange potential and one $D^{(*)}$ meson exchange potential from the above interaction Lagrangians.
We introduce the monopole-type form factor,
\begin{align}
F(\vec{q}) = \frac{\Lambda^2-m_a^2}{\Lambda^2+|\vec{q}|^2}~,
\end{align}
at each vertex,  where 
$\Lambda$ is a cutoff parameter, 
$m_a$ and $\vec{q}$ are the mass and momentum of exchanging particle, respectively.  
Although the 
cutoff parameter $\Lambda$ may be different for pion and $D^{(*)}$ meson,  
we use the same value in the present analysis   for simplicity.
Including this form factor, the exchange potentials are written as
\begin{align}
V_{ij}^{a}(r) &= G_{ij}C_a(r, \Lambda, m_a)~, 
\label{pot part}
\end{align}
where $G_{ij}$ denotes the coefficients, coupling constants, spin factors, and isospin factors for each $(i, j)$ channel.
$C_a(r, \Lambda, m_a)$ is defined as
\begin{align}
C_a(r, \Lambda, m_a) &= \frac{m_a^2}{4\pi}\left[ \frac{e^{-m_ar} - e^{-\Lambda r} }{r} - \frac{\Lambda^2-m_a^2}{2\Lambda}e^{-\Lambda r} \right]~.
\end{align}
The explicit forms of potential are shown in the following sections.

%%%%%%%%%%%%%%%%%%%%%%%%%%%%%%%%%%%%%%%%%%%%%%%%%%%%%%%%%%

\section{Numerical result for pentaquark $P_c(4380)$}
\label{sec:pc4380}
We consider the $\Jp N - \Lambda_c\bar{D}^* - \Sigma_c^*\bar{D} - \Sigma_c\bar{D}^* - \Sigma_c^*\bar{D}^*$ coupled system with $S$-wave orbital angular momentum.
We solve the coupled channel Schr{\"o}dinger equation, using the
potential $V(r)$ given by a $5\times5$ matrix expressed as
\begin{widetext}
\begin{align}
V(r) &= \left(
\begin{array}{ccccc}
0 & G_1G_3(C_D + C_{D^*}) & -2\sqrt{6}G_1G_2C_{D^*} & \sqrt{2}G_1G_2(3C_D-C_{D^*}) & 2\sqrt{10}G_1G_2C_{D^*} \vspace{2pt} \\
G_1G_3(C_D + C_{D^*}) & 0 & -\frac{gg_4}{\sqrt{6}\fpi^2}C_{\pi} & \frac{gg_4}{3\sqrt{2}\fpi^2}C_{\pi} & -\frac{\sqrt{10}gg_4}{6\fpi^2}C_{\pi} \vspace{2pt} \\
-2\sqrt{6}G_1G_2C_{D^*} & -\frac{gg_4}{\sqrt{6}\fpi^2}C_{\pi} & 0 & \frac{gg_1}{2\sqrt{3}\fpi^2}C_{\pi} & -\frac{\sqrt{15}gg_1}{9\fpi^2}C_{\pi} \vspace{2pt} \\
\sqrt{2}G_1G_2(3C_D-C_{D^*}) & \frac{gg_4}{3\sqrt{2}\fpi^2}C_{\pi} & \frac{gg_1}{2\sqrt{3}\fpi^2}C_{\pi} & -\frac{gg_1}{3\fpi^2}C_{\pi} & \frac{\sqrt{5}gg_1}{6\fpi^2}C_{\pi} \vspace{2pt} \\
2\sqrt{10}G_1G_2C_{D^*} & -\frac{\sqrt{10}gg_4}{6\fpi^2}C_{\pi} & -\frac{\sqrt{15}gg_1}{9\fpi^2}C_{\pi} & \frac{\sqrt{5}gg_1}{6\fpi^2}C_{\pi} & -\frac{2gg_1}{9\fpi^2}C_{\pi}
\end{array}
\right)~,
\label{eq:pot5by5}
\end{align}
\end{widetext}
where $C_a$ is defined in Eq.~(\ref{pot part}).
The wave function has five components;
\begin{align}
\Psi(r) = \left(
\begin{array}{c}
\psi_{\Jp N} \\ \psi_{\Lamc\Dbs} \\ \psi_{\Sigcs\Db} \\ \psi_{\Sigc\Dbs} \\ \psi_{\Sigcs\Dbs}
\end{array}
\right)~.
\end{align}
We use $m_{\pi}=137.2$, $m_{N}=938.9$, $m_{D}=1867.2$, $m_{D^*}=2008.6$, $m_{\Lamc}=2286.5$, $m_{\Sigc}=2453.5$, $m_{\Sigcs}=2518.1$ and $m_{\Jp}=3096.9$ MeV for the hadron masses \cite{Olive:2016xmw}.
The thresholds for the hadronic molecules are shown in Table \ref{tab:pc4380}.
In this calculation, we vary the cutoff parameter $\Lambda$ from $1000$ to $1500$ MeV.
For the coupling constant $g_1$, we use $g_1=0.942$ estimated in a quark model \cite{Liu:2001ce}
as a reference value,  and study the $g_1$ dependence of the results using $g_1=0.753$ and $1.13$.  
To obtain the bound and resonance solutions, we use the complex scaling method \cite{Aguilar:1971ve, Balslev:1971vb, Aoyama:2006csm} and Gaussian expansion method \cite{Hiyama:2003cu, Hiyama:2012sma}.

The  resultant  complex energies are shown in Table \ref{tab:pc4380}.
When the cutoff parameter $\Lambda$ becomes larger, the mass and width become smaller.
In our ranges of $\Lambda$ and $g_1$, the bound state solution which has the real energy below the $\Jp N$ threshold does not appear.
The solutions of $\Lambda=1200$ and 1300 MeV for $g_1=0.942$ can reproduce the observed  mass of $P_c(4380)$, $4380\pm8\pm29$MeV and width,  $205\pm18\pm86$\,MeV. 
However,  there exists another resonance state solution, the mass of which is $4283.1$MeV for $\Lambda = 1200$\,MeV and $4227.1$MeV for $\Lambda = 1300$\,MeV. 
These lower states are not observed in LHCb experiment, therefore we consider that these parameter sets are unlikely.
On the other hand, for the $\Lambda=1000$MeV and $g_1=0.753$, we obtain only one resonance state which corresponds to $P_c(4380)$.
Its mass, $4390.2$\,MeV, is slightly above the $\Sigma_c^*\bar{D}$ threshold, so this state is
interpreted as a resonance state of $\Sigma_c^*\bar{D}$ molecule.

\begin{table*} [!htbp]
\centering
\caption{
 Energy eigen values in 
$\Jp N - \Lambda_c\bar{D}^* - \Sigma_c^*\bar{D} - \Sigma_c\bar{D}^* - \Sigma_c^*\bar{D}^*$ coupled system with $S$-wave states in $J^P=3/2^-$.
We show the thresholds of each hadronic molecular state in the last line of the table for a reference. 
}
\begin{tabular}{c|cccccc}\hline
 & & & $\Lambda$ [MeV] & & & \\ \hline
$g_1$ & $1000$ & $1100$ & $1200$ & $1300$ & $1400$ & $1500$ \\ \hline
\shortstack{$0.753$ \\ $$} & \shortstack{\\$---$ \\ $4390.2-i109$} & \shortstack{\\$---$ \\ $4352.5-i61.4$} & \shortstack{\\$---$ \\ $4312.2-i36.5$} & \shortstack{\\$4449.4-i172$ \\ $4277.9-i8.1$} & \shortstack{\\$4397.1-i80.8$ \\ $4208.9-i5.0$} & \shortstack{\\$4344.0-i31.4$ \\ $4159.5-i2.2$} \\ \hline
\shortstack{$0.942$ \\ $$} & \shortstack{\\$---$ \\ $4333.6-i66.4$} & \shortstack{\\$4438.5-i143$ \\ $4308.5-i36.3$} & \shortstack{\\$4416.9-i126$ \\ $4283.1-i12.1$} & \shortstack{\\$4397.2-i100$ \\ $4227.1-i8.9$} & \shortstack{\\$4345.0-i51.5$ \\ $4190.6-i6.22$} & \shortstack{\\$4314.3-i21.0$ \\ $4150.0-i5.2$} \\ \hline
\shortstack{$1.13$ \\ $$} & \shortstack{\\$4422.8-i99.6$ \\ $4315.5-i36.8$} & \shortstack{\\$4382.5-i70.9$ \\ $4273.2-i19.7$} & \shortstack{\\$4359.6-i53.2$ \\ $4237.6-i10.9$} & \shortstack{\\$4295.7-i26.3$ \\ $4187.8-i6.58$} & \shortstack{\\$4226.6-i7.19$ \\ $4126.8-i3.00$} & \shortstack{\\$4149.6-i3.20$ \\ $4051.1-i0.578$} \\ \hline\hline
 & threshold[MeV] & \hspace{2pt}$\Jp N$(4035.8) \hspace{9pt} & $\Lamc\Dbs$(4295.1)\hspace{9pt} & $\Sigcs\Db$(4385.3)\hspace{9pt} & $\Sigc\Dbs$(4462.1)\hspace{9pt} & $\Sigcs\Dbs$(4526.7)\hspace{9pt} \\ \hline
\end{tabular}
\label{tab:pc4380}
\end{table*}

%%%%%%%%%%%%%%%%%%%%%%%%%%%%%%%%%%%%%%%%%%%%%%%%%%%%%%%%%%%%%

\section{Doubly charmed baryon $\Xi_{cc}^*(4380)$}
\label{sec:xicc4380}

We study the doubly charmed baryon as a hadronic molecular state in this section.
 Replacing  $\bar{D}^{(*)}$ with $D^{(*)}$ and excluding the $\Jp N$ channel from the calculation in Sec.~\ref{sec:pc4380}, 
we construct the $ccqq\bar{q}$ state which has the same flavor quantum number as the $ccq$ baryon has.
The interactions of one-pion exchange is not changed 
by the replacement of $D^{(*)}$ meson.
Therefore, the corresponding potential matrix is a bottom-right 4$\times$4 block of  
Eq.~(\ref{eq:pot5by5}): 
\begin{align}
V(r) &= \left(
\begin{array}{cccc}
0 & -\frac{gg_4}{\sqrt{6}\fpi^2} & \frac{gg_4}{3\sqrt{2}\fpi^2} & -\frac{\sqrt{10}gg_4}{6\fpi^2} \vspace{2pt} \\
-\frac{gg_4}{\sqrt{6}\fpi^2} & 0 & \frac{gg_1}{2\sqrt{3}\fpi^2} & -\frac{\sqrt{15}gg_1}{9\fpi^2} \vspace{2pt} \\
\frac{gg_4}{3\sqrt{2}\fpi^2} & \frac{gg_1}{2\sqrt{3}\fpi^2} & -\frac{gg_1}{3\fpi^2} & \frac{\sqrt{5}gg_1}{6\fpi^2} \vspace{2pt} \\
-\frac{\sqrt{10}gg_4}{6\fpi^2} & -\frac{\sqrt{15}gg_1}{9\fpi^2} & \frac{\sqrt{5}gg_1}{6\fpi^2} & -\frac{2gg_1}{9\fpi^2}
\end{array}
\right)C_{\pi}~.
\label{eq:pot4by4}
\end{align}
The wave function has four components;
\begin{align}
\Psi(r) = \left(
\begin{array}{c}
\psi_{\Lamc\Dbs} \\ \psi_{\Sigcs\Db} \\ \psi_{\Sigc\Dbs} \\ \psi_{\Sigcs\Dbs}
\end{array}
\right)~.
\end{align}

We investigate the dependence on  the cutoff $\Lambda$ and coupling constant $g_1$ in the same range as in Sec.\ref{sec:pc4380},
 and 
show the numerical results 
in Table~\ref{tab:Xi4380}.
\begin{table*} [!htbp]
\centering
\caption{
Energy eigenvalues in 
$\Lambda_c D^* - \Sigma_c^*D - \Sigma_cD^* - \Sigma_c^*D^*$ coupled system with $S$-wave states in $J^P=3/2^-$.
We show the thresholds of each hadronic molecular state 
at the last line of the table for a reference. 
}
\begin{tabular}{c|cccccc}\hline
 & & & $\Lambda$ [MeV] & & & \\ \hline
$g_1$ & $1000$ & $1100$ & $1200$ & $1300$ & $1400$ & $1500$ \\ \hline
\shortstack{$0.753$ \\ $$} & \shortstack{\\$---$ \\ $4370.1-i68.7$} & \shortstack{\\$---$ \\ $4340.9-i55.6$} & \shortstack{\\$4440.9-i120$ \\ $4302.4-i15.7$} & \shortstack{\\$4420.9-i99.6$ \\ $4262.3$} & \shortstack{\\$4386.8-i73.7$ \\ $4214.3$} & \shortstack{\\$4347.4-i25.2$ \\ $4166.0$} \\ \hline
\shortstack{$0.942$ \\ $$} & \shortstack{\\$---$ \\ $4350.3-i69.1$} & \shortstack{\\$4448.9-i142$ \\ $4325.5-i31.3$} & \shortstack{\\$4424.5-i122$ \\ $4290.4$} & \shortstack{\\$4401.6-i98.7$ \\ $4242.3$} & \shortstack{\\$4367.0-i70.4$ \\ $4200.1$} & \shortstack{\\$4328.4-i29.7$ \\ $4167.8$} \\ \hline
\shortstack{$1.13$ \\ $$} & \shortstack{\\$4414.0-i86.0$ \\ $4325.4-i11.3$} & \shortstack{\\$4377.1-i73.8$ \\ $4295.3-i0.1$} & \shortstack{\\$4342.4-i27.9$ \\ $4265.1$} & \shortstack{\\$4296.7-i0.2$ \\ $4226.5$} & \shortstack{\\$4247.8$ \\ $4180.8$} & \shortstack{\\$4185.4$ \\ $4117.9$} \\ \hline\hline
 & threshold[MeV] & \hspace{2pt}$\Lamc D^*$(4295.1)\hspace{9pt} & $\Sigcs D$(4385.3)\hspace{9pt} & $\Sigc D^*$(4462.1)\hspace{9pt} & $\Sigcs D^*$(4526.7)\hspace{9pt} \\ \hline
\end{tabular}
\label{tab:Xi4380}
\end{table*}
Comparing the results of Table~\ref{tab:pc4380} and Table \ref{tab:Xi4380}, they have close mass and decay width.
For $\Lambda = 1200$-$1500$\,MeV,
we obtain bound state solutions whose masses are below the threshold of $\Lambda_cD^*$.
Since the mass and width of $P_c(4380)$ are not within experimental errors for $\Lambda \ge 1100$\,MeV,~\footnote{
As we stated in the previous section, there are a few parameter choices for which the mass and width of $P_c(4380)$ are reproduced even for $\Lambda \ge 1100$\,MeV.  However, there is another state lighter than $P_c(4380)$, so that these parameter choices are unlikely. 
}
 the bound state below $\Lambda_c D^\ast$ is unlikely to exist.
On the other hand, when $\Lambda = 1000$\,MeV and $g_1 = 0.753$ are used, for which the mass and width of $P_c(4380)$ are within experimental errors, the mass and width of the doubly charmed baryon are $M= 4370.1$\,MeV and $\Gamma = 68.7$\,MeV, which are close to those of $P_c(4380)$.
This means that, when the hidden charm pentaquark $P_c(4380)$ exist as a hadronic molecular state, 
a 
doubly charmed baryon 
with same spin and parity exists, and its mass and width are close to $P_c(4380)$, which  
we call this doubly charmed baryon $\Xi_{cc}^*(4380)$.

%%%%%%%%%%%%%%%%%%%%%%%%%%%%%%%%%%%%%%%%%%%%%%%%%%%%%

\section{Summary and Discussions}
\label{sec:summary}

We 
investigated the coupled channel of the $\Jp N - \Lambda_c\bar{D}^* - \Sigma_c^*\bar{D} - \Sigma_c\bar{D}^* - \Sigma_c^*\bar{D}^*$ in $J^P=3/2^-$ with $S$-wave orbital angular momentum.
We constructed the one-pion exchange and one-$D^{(*)}$ meson exchange potential and solved the complex scaled Schr{\"o}dinger-type equation.
We showed that, 
for $\Lambda= 1200$-$1300$\,MeV, there exists another state having mass and with smaller than $P_c(4380)$, while 
for $\Lambda = 1000$\,MeV and $g_1 = 0.753$, there exists only one molecular state having the mass and width within errors of experimental values.
This shows that hidden charm pentaquark $P_c(4380)$ can be explained  as a  
$S$-wave hadronic molecular state.

We 
studied the coupled channel of the $\Lambda_cD^* - \Sigma_c^*D - \Sigma_cD^* - \Sigma_c^*D^*$ in $J^P=3/2^-$ with $S$-wave orbital angular momentum.
Since the 
one-pion  interactions for $\bar{D}^{(*)}$ mesons are the same as the ones for $D^{(*)}$ mesons, we obtain a $\Xi_{cc}$ state with $J^P= \frac{3}{2}^-$ as a resonance state whose mass and width are very close to those of $P_c(4380)$, which we call $\Xi_{cc}^{\ast}(4380)$.

We think that the same mechanism applies for $P_c(4450)$: When $P_c(4450)$ is described as
a hadronic molecular state, 
there exists a doubly charmed baryon which has 
a mass and a width quite close to $P_c(4450)$.

Although we do not evaluate 
the partial decay width for $J/\psi N$ in this paper, 
we can see that the partial width is much narrower than that for $\Lambda_c\bar{D}^*$ in the following way: 
When we omit the contribution from $J/\psi N$ channel to $P_c(4380)$, 
the relevant potential become the same as that for $\Xi_{cc}^*(4380)$ . 
This implies that the resultant mass and width without $J/\psi N$ channel is already close to the ones with $J/\psi N$ channel. 
This is consistent with the analysis of decay behaviors in Ref.~\cite{Lin:2017mtz}.

Comparing the results of $P_c(4380)$ and $\Xi_{cc}^*(4380)$, we can see that the contribution of the $J/\psi N$ channel to $P_c(4380)$ is small.
This is consistent with 
the naive prospect of the supression of $D^{(*)}$ meson exchange potentials.
Our  evaluation 
of the coupling to the $J/\psi N$ was very rough, 
 so that the values used in this analysis include some ambiguities.  
Furthermore, there may exist other contributions which couple the $J/\psi N$ channel to
$ \Lambda_c\bar{D}^* - \Sigma_c^*\bar{D} - \Sigma_c\bar{D}^* - \Sigma_c^*\bar{D}^*$.
 We think that these ambiguities do not change our results, since the contribution from $J/\psi N$ channel is very small consistently with the result in Ref.~\cite{Lin:2017mtz}.

In the present analysis, we 
do not include the decay of $\Sigma_c^* \to \Lambda_c\pi$ for $\Sigma_c^*\bar{D}^{(*)}$ state.
The width of this decay is about 15MeV \cite{Olive:2016xmw}, 
so it makes the total width of $P_c(4380)$ broader~\cite{Lin:2017mtz}.

We used only one-pion exchange potential for $\Lambda_c{D}^{(*)} - \Sigma_c^{(*)}{D}^{(*)}$ coupled channel in the analysis of $\Xi_{cc}^*(4380)$, 
which is the same as the one for $\Lambda_c\bar{D}^{(*)} - \Sigma_c^{(*)}\bar{D}^{(*)}$ coupled channel in the analysis of $P_c(4380)$.
Then, we obtained the mass and width of $\Xi_{cc}^*(4380)$ very close to those of $P_c(4380)$.
When we include the effects of $\omega$ meson exchange, difference between $DD\omega$ and $D\bar{D}\omega$ will generate some differences of the mass and width~\cite{Chen:2017vai}.

There are some theoretical predictions of ordinary $ccq$-type baryons in $J^P=3/2^-$ 
\cite{Migura:2006ep, Chiu:2005zc, Wang:2010it, Karliner:2014gca, Padmanath:2015jea, Wei:2015gsa, Shah:2017liu}.
In Ref.~\cite{Shah:2017liu}, the mass of $3P$-state spin-$\frac{3}{2}$ $\Xi_{cc}$ is predicted to be about $4.41$\,GeV. This state might mix with $\Xi_{cc}^*(4380)$ predicted in this analysis.

We expect that the precise properties of $P_c$ pentaquarks and the existence of excited $\Xi_{cc}$ baryons would be revealed in future experiments.

%%%%%%%%%%%%%%%%%%%%%%%%%%

\acknowledgments
We would like to thank Yuji Kato for useful discussion.
The work of Y.S. is supported in part by JSPS Grant-in-Aid for JSPS Research Fellow No. JP17J06300. 
The work of M.H. is supported in part by 
the JSPS Grant-in-Aid for Scientific Research (C) No.~16K05345.

\end{document}